\documentclass[twocolumn,tighten]{aastex62}
\newcommand{\target}{TOI-1130$ $}
\newcommand{\thisstar}{TOI-1130}
\newcommand{\thisplanetb}{TOI-1130~b$ $}
\newcommand{\thisplanetc}{TOI-1130~c$ $}

\newcommand{\rsun}{\ensuremath{R_\sun}}
\newcommand{\msun}{\ensuremath{M_\sun}}
\newcommand{\lsun}{\ensuremath{L_\sun}}

\newcommand{\rearth}{\ensuremath{R_\earth}}
\newcommand{\mearth}{\ensuremath{M_\earth}}

\newcommand{\rjup}{\ensuremath{R_{\rm J}}}
\newcommand{\mjup}{\ensuremath{M_{\rm J}}}

\newcommand{\teff}{\ensuremath{T_{\rm eff}}}

\newcommand{\rpl}{\ensuremath{R_{p}}}
\newcommand{\mpl}{\ensuremath{M_{p}}}

\newcommand{\rhopl}{\ensuremath{\rho_{p}}}

\newcommand{\rstar}{\ensuremath{R_\star}}
\newcommand{\mstar}{\ensuremath{M_\star}}
\newcommand{\lstar}{\ensuremath{L_\star}}

\newcommand{\teffstar}{\ensuremath{T_{\rm eff}}}
\newcommand{\rhostar}{\ensuremath{\rho_\star}}
\newcommand{\loggstar}{\ensuremath{\log{g}}}

\newcommand{\arstar}{\ensuremath{a/\rstar}}

\newcommand{\kms}{km~s$^{-1}$}

\newcommand{\TESS}{\emph{TESS}}

\newcommand{\gaia}{\emph{Gaia}}

\newcommand{\gcmc}{\ensuremath{\rm g\,cm^{-3}}}
\newcommand{\ergscmsq}{\ensuremath{\rm erg\,s^{-1}\,cm^{-2}}}

\newcommand{\tessfitPb}{\ensuremath{4.066499_{-0.000045}^{+0.000046}}}
\newcommand{\tessfitPc}{\ensuremath{8.350381_{-0.000033}^{+0.000032}}}
\newcommand{\tessfitTcc}{\ensuremath{2458657.90461_{-0.00022}^{+0.00021}}}
\newcommand{\tessfitTcb}{\ensuremath{2458658.74627_{-0.00068}^{+0.00072}}}
\newcommand{\tessfitRratiob}{\ensuremath{0.04860_{-0.00090}^{+0.00111}}}
\newcommand{\tessfitRratioc}{\ensuremath{0.218_{-0.029}^{+0.037}}}
\newcommand{\tessfitbb}{\ensuremath{0.48_{-0.21}^{+0.11}}}
\newcommand{\tessfitbc}{\ensuremath{0.995_{-0.043}^{+0.046}}}
\newcommand{\tessfitKc}{\ensuremath{0.1259_{-0.0055}^{+0.0052}}}
\newcommand{\tessfitesinwb}{\ensuremath{0.33_{-0.33}^{+0.19}}}
\newcommand{\tessfitecoswb}{\ensuremath{0.26_{-0.17}^{+0.13}}}
\newcommand{\tessfitesinwc}{\ensuremath{0.17_{-0.16}^{+0.11}}}
\newcommand{\tessfitecoswc}{\ensuremath{-0.093_{-0.072}^{+0.079}}}
\newcommand{\tessfitmstar}{\ensuremath{0.684_{-0.017}^{+0.016}}}
\newcommand{\tessfitrstar}{\ensuremath{0.687_{-0.015}^{+0.015}}}
\newcommand{\tessfitloggstar}{\ensuremath{4.60_{-0.018}^{+0.02}}}
\newcommand{\tessfitrhostar}{\ensuremath{2.97_{-0.17}^{+0.20}}}
\newcommand{\tessfitsemib}{\ensuremath{0.04394_{-0.00038}^{+0.00035}}}
\newcommand{\tessfitsemic}{\ensuremath{0.07098_{-0.00060}^{+0.00056}}}
\newcommand{\tessfitaorb}{\ensuremath{13.75_{-0.27}^{+0.31}}}
\newcommand{\tessfitaorc}{\ensuremath{22.21_{-0.43}^{+0.50}}}
\newcommand{\tessfitincb}{\ensuremath{87.98_{-0.46}^{+0.86}}}
\newcommand{\tessfitincc}{\ensuremath{87.43_{-0.16}^{+0.16}}}
\newcommand{\tessfitRpb}{\ensuremath{3.65_{-0.10}^{+0.10}}}
\newcommand{\tessfitRpc}{\ensuremath{1.50_{-0.22}^{+0.27}}}
\newcommand{\tessfitec}{\ensuremath{0.047_{-0.027}^{+0.040}}}
\newcommand{\tessfiteb}{\ensuremath{0.22_{-0.11}^{+0.11}}}
\newcommand{\tessfitMpc}{\ensuremath{0.974_{-0.044}^{+0.043}}}
\newcommand{\tessfitwc}{\ensuremath{-28_{-55}^{+24}}}

\newcommand{\tessfitrhoc}{\ensuremath{0.38_{-0.15}^{+0.24}}}

\newcommand{\tessfitTdurb}{\ensuremath{2.30_{-0.14}^{+0.18}}}
\newcommand{\tessfitTdurc}{\ensuremath{2.02_{-0.044}^{+0.044}}}

\newcommand{\tessfitlumstar}{\ensuremath{0.140_{-0.010}^{+0.011}}}
\newcommand{\tessfitTeqb}{\ensuremath{810_{-15}^{+15}}}
\newcommand{\tessfitTeqc}{\ensuremath{637_{-12}^{+12}}}
\newcommand{\chironfitgamma}{\ensuremath{-9.5241_{-0.0040}^{+0.0041}}}
\newcommand{\chironfitjit}{\ensuremath{0.0112_{-0.0069}^{+0.0066}}}

\newcommand\mysim{\mathord{\sim}}

\shorttitle{A new hot Jupiter with a neighbor}
\shortauthors{}

\begin{document}

\title{TESS spots a hot Jupiter with an inner transiting Neptune}

\author[0000-0003-0918-7484]{Chelsea X.\ Huang}
\altaffiliation{Juan Carlos Torres Fellow}
\affiliation{Department of Physics, and Kavli Institute for Astrophysics and Space Research, Massachusetts Institute of Technology, Cambridge, MA 02139, USA}

\author[0000-0002-8964-8377]{Samuel N. Quinn}
\affiliation{Center for Astrophysics \textbar \ Harvard \& Smithsonian, 60 Garden Street, Cambridge, MA 02138, USA}

\author[0000-0001-7246-5438]{Andrew Vanderburg}
\altaffiliation{NASA Sagan Fellow}
\affiliation{Department of Astronomy, The University of Texas at Austin, Austin, TX 78712, USA}

\author[0000-0002-7733-4522]{Juliette Becker}
\altaffiliation{51 Pegasi b Fellow}
\affiliation{Division of Geological and Planetary Sciences, California Institute of Technology, Pasadena, CA 91125, USA}

\author[0000-0001-8812-0565]{Joseph E. Rodriguez}
\affiliation{Center for Astrophysics \textbar \ Harvard \& Smithsonian, 60 Garden Street, Cambridge, MA 02138, USA}

\author[0000-0003-1572-7707]{Francisco J. Pozuelos}
\affiliation{Space Sciences, Technologies and Astrophysics Research (STAR) Institute, Université de Liège, 19C Allée du 6 Août, 4000 Liège, Belgium}
\affiliation{Astrobiology Research Unit, Université de Liège, 19C Allée du 6 Août, 4000 Liège, Belgium}

\author[0000-0001-8627-9628]{Davide Gandolfi}
\affiliation{Dipartimento di Fisica, Universit\`a degli Studi di Torino, via Pietro Giuria 1, I-10125, Torino, Italy.}

\author{George Zhou}
\altaffiliation{NASA Hubble Fellow}
\affiliation{Center for Astrophysics \textbar \ Harvard \& Smithsonian, 60 Garden Street, Cambridge, MA 02138, USA}

\author[0000-0003-3654-1602]{Andrew W. Mann}
\affiliation{Department of Physics and Astronomy, University of North Carolina at Chapel Hill, Chapel Hill, NC 27599, USA}

\author[0000-0001-6588-9574]{Karen A.\ Collins}
\affiliation{Center for Astrophysics \textbar \ Harvard \& Smithsonian, 60 Garden Street, Cambridge, MA 02138, USA}

\author{Ian Crossfield}
\affiliation{Department of Physics and Astronomy, University of Kansas, 1251 Wescoe Hall Dr., Lawrence, KS 66045, USA}
\affiliation{Department of Physics and Kavli Institute for Astrophysics and Space Research, Massachusetts Institute of Technology, Cambridge, MA 02139, USA}

\author{Khalid Barkaoui} 
\affiliation{Astrobiology Research Unit, Université de Liège, 19C Allée du 6 Août, 4000 Liège, Belgium} 
\affiliation{Oukaimeden Observatory, High Energy Physics and Astrophysics Laboratory, Cadi Ayyad University, Marrakech, Morocco}

\author[0000-0003-2781-3207]{Kevin I.\ Collins}
\affiliation{George Mason University, 4400 University Drive, Fairfax, VA, 22030 USA}

\author[0000-0002-0855-8426]{Malcolm Fridlund}
\affiliation{Department of Space, Earth and Environment, Chalmers University of Technology, Onsala Space Observatory, SE-439 92 Onsala, Sweden}
\affiliation{Leiden Observatory, University of Leiden, Leiden, The Netherlands}

\author[0000-0003-1462-7739]{Micha{\"e}l Gillon} 
\affiliation{Astrobiology Research Unit, Université de Liège, 19C Allée du 6 Août, 4000 Liège, Belgium}

\author{Erica J. Gonzales}
\altaffiliation{NSF Graduate Research Fellow}
\affiliation{Department of Astronomy and Astrophysics University of California, Santa Cruz 
,1156 High St, Santa Cruz California 95064}

\author[0000-0002-3164-9086]{Maximilian N.\ G{\"u}nther}
\altaffiliation{Juan Carlos Torres Fellow}
\affiliation{Department of Physics, and Kavli Institute for Astrophysics and Space Research, Massachusetts Institute of Technology, Cambridge, MA 02139, USA}

\author{Todd J. Henry}        
\affiliation{RECONS Institute, Chambersburg, PA 17201, USA}

\author{Steve, B. Howell}
\affiliation{Space Science and Astrobiology Division,
NASA Ames Research Center M/S 245-6, Moffett Field, CA 94035}

\author{Hodari-Sadiki James}  
\affiliation{Georgia State University, Atlanta, GA 30302, USA}

\author{Wei-Chun Jao}         
\affiliation{Georgia State University, Atlanta, GA 30302, USA}

\author{Emmanu{\"e}l Jehin}
\affiliation{Space Sciences, Technologies and Astrophysics Research (STAR) Institute, Université de Liège, 19C Allée du 6 Août, 4000 Liège, Belgium}

\author[0000-0002-4625-7333]{Eric L. N. Jensen}
\affiliation{Dept.\ of Physics \& Astronomy, Swarthmore College, Swarthmore PA 19081, USA}

\author[0000-0002-7084-0529]{Stephen R. Kane}
\affiliation{Department of Earth and Planetary Sciences, University of California,
Riverside, CA 92521, USA}

\author{Jack J. Lissauer}\affiliation{Space Science \& Astrobiology Division, NASA Ames Research Center, Moffett Field, CA 94035, USA}

\author{Elisabeth Matthews}\affiliation{Department of Physics, and Kavli Institute for Astrophysics and Space Research, Massachusetts Institute of Technology, Cambridge, MA 02139, USA}

\author[0000-0001-7233-7508]{Rachel A. Matson}
\affiliation{U.S. Naval Observatory, Washington, DC 20392, USA}

\author{Leonardo A. Paredes}
\affiliation{Georgia State University, Atlanta, GA 30302, USA}

\author{Joshua E. Schlieder} 
\affiliation{NASA Goddard Space Flight Center, 8800 Greenbelt Road, Greenbelt, MD 20771, USA}

\author[0000-0002-3481-9052]{Keivan G.\ Stassun}
\affiliation{Vanderbilt University, Department of Physics \& Astronomy, 6301 Stevenson Center Lane, Nashville, TN 37235, USA}
\affiliation{Fisk University, Department of Physics, 1000 18th Ave.\ N., Nashville, TN 37208, USA}

\author[0000-0002-1836-3120]{Avi Shporer}
\affiliation{Department of Physics, and Kavli Institute for Astrophysics and Space Research, Massachusetts Institute of Technology, Cambridge, MA 02139, USA}

\author[0000-0001-5401-8079]{Lizhou Sha}
\affiliation{Department of Physics, and Kavli Institute for Astrophysics and Space Research, Massachusetts Institute of Technology, Cambridge, MA 02139, USA}

\author[0000-0001-5603-6895]{Thiam-Guan Tan}
\affiliation{Perth Exoplanet Survey Telescope, Perth, Western Australia}

\author{Iskra Georgieva}
\affiliation{Department of Space, Earth and Environment, Chalmers University of Technology, Onsala Space Observatory, SE-439 92 Onsala, Sweden}

\author{Savita Mathur}
\affiliation{Instituto de Astrofisica de Canarias
c/ Via Lactea s/n,
38205 La Laguna
Santa Cruz de Tenerife
SPAIN}

\author{Enric Palle}
\affiliation{Instituto de Astrofisica de Canarias,
Via Lactea sn, 38200, La Laguna, Tenerife, Spain}

\author{Carina M. Persson}
\affiliation{Department of Space, Earth and Environment, Chalmers University of Technology, Onsala Space Observatory, SE-439 92 Onsala, Sweden}

\author{Vincent Van Eylen}
\affiliation{Mullard Space Science Laboratory, Department of Space and Climate Physics, University College London}

\author{George R.\ Ricker}
\affiliation{Department of Physics, and Kavli Institute for Astrophysics and Space Research, Massachusetts Institute of Technology, Cambridge, MA 02139, USA}

\author[0000-0001-6763-6562]{Roland K. Vanderspek}
\affiliation{Department of Physics, and Kavli Institute for Astrophysics and Space Research, Massachusetts Institute of Technology, Cambridge, MA 02139, USA}

\author[0000-0001-9911-7388]{David W.\ Latham}
\affiliation{Center for Astrophysics \textbar \ Harvard \& Smithsonian, 60 Garden Street, Cambridge, MA 02138, USA}

\author[0000-0002-4265-047X]{Joshua N.\ Winn}
\affiliation{Department of Astrophysical Sciences, Princeton University, 4 Ivy Lane, Princeton, NJ 08544, USA}

\author[0000-0002-6892-6948]{S.~Seager}
\affiliation{Department of Physics and Kavli Institute for Astrophysics and Space Research, Massachusetts Institute of Technology, Cambridge, MA 02139, USA}
\affiliation{Department of Earth, Atmospheric and Planetary Sciences, Massachusetts Institute of Technology, Cambridge, MA 02139, USA}
\affiliation{Department of Aeronautics and Astronautics, MIT, 77 Massachusetts Avenue, Cambridge, MA 02139, USA}

\author[0000-0002-4715-9460]{Jon M. Jenkins}
\affiliation{NASA Ames Research Center, Moffett Field, CA 94035, USA}

\author[0000-0002-7754-9486]{Christopher~J.~Burke}
\affiliation{Kavli Institute for Astrophysics and Space Research, Massachusetts Institute of Technology, Cambridge, MA, USA}

\author{Robert F. Goeke}
\affiliation{Department of Physics, and Kavli Institute for Astrophysics and Space Research, Massachusetts Institute of Technology, Cambridge, MA 02139, USA}

\author{Stephen Rinehart}
\affiliation{NASA Goddard Space Flight Center, Greenbelt, MD,
USA.}

\author[0000-0003-4724-745X]{Mark E. Rose}
\affiliation{NASA Ames Research Center, Moffett Field, CA 94035, USA}

\author[0000-0002-8219-9505 ]{Eric B. Ting}
\affiliation{NASA Ames Research Center, Moffett Field, CA 94035, USA}

\author[0000-0002-5286-0251]{Guillermo Torres}
\affiliation{Center for Astrophysics \textbar \ Harvard \& Smithsonian, 60 Garden Street, Cambridge, MA 02138, USA}

\author[0000-0001-9665-8429]{Ian~Wong}
\altaffiliation{51 Pegasi b Fellow}
\affiliation{Department of Earth, Atmospheric and Planetary Sciences, Massachusetts Institute of Technology,
Cambridge, MA 02139, USA}

\begin{abstract}

Hot Jupiters are rarely accompanied by other planets within a factor of a few in orbital distance. Previously, only two such systems have been found. Here, we report the discovery of a third system using data from
the \textit{Transiting Exoplanet Survey Satellite} (\TESS). The host star, \target,
is an 11th magnitude K-dwarf in Gaia G band. It has two transiting planets:
a Neptune-sized planet ($3.65\pm 0.10$\,\rearth) with a 4.1-day period, and a hot Jupiter
($1.50^{+0.27}_{-0.22}$\,\rjup\,) with an 8.4-day period. Precise radial-velocity
observations show that the mass of the hot Jupiter is $0.974^{+0.043}_{-0.044}$\,\mjup.
For the inner Neptune, the data provide only an upper limit on the mass
of 0.17\,\mjup~(3$\sigma$).  Nevertheless, we are confident the inner
planet is
real, based on follow-up ground-based photometry and adaptive optics imaging that rule
out other plausible sources of the \TESS\ transit signal.
The unusual planetary architecture of and the brightness of the host star
make \target\  a good test case for planet formation theories, and an attractive
target for future spectroscopic observations.  
\end{abstract}


\section{Introduction}
\label{sec:intro}

The origin of gas giants on extremely short-period orbits has been an unsolved problem for 25 years \citep{MayorQueloz1995}. Although many scenarios have been proposed to place these hot Jupiters in their current orbital locations (disk migration, \emph{in situ} formation, planet-planet scattering, secular migration, etc.), no single mechanism seems
capable of satisfying all the observational constraints \citep{DawsonJohnson2018}. One clue
is that hot Jupiters tend to be ``lonely,'' in the sense that stars with hot Jupiters
often have wide-orbiting companions \citep{SchlaufmanWinn2016} but tend to lack nearby planetary companions within a factor of 2 or 3 in orbital distance \citep{Steffen+2012}.
The only known exceptions are WASP-47 and Kepler-730 \citep{Becker:2015b, zhu:2018, Canas:2019}.

How should these two systems be understood? Are they simply
rare variants of hot Jupiters?  Or did they form by a different process --- 
perhaps the same process that led to the formation of ``warm Jupiters''
($P=$~20 to 100 days), which are often flanked by smaller companions \citep{Huang:2016}? The Transiting Exoplanet Survey Satellite \citep[\TESS{};][]{ricker} is well suited to address these questions by discovering more systems like WASP-47 and Kepler-730. By observing
most of the sky, \TESS{} is expected to find thousands of hot Jupiters \citep{Sullivan2015}, while also
having good enough photometric precision to find smaller planets around the same
stars \citep[see, e.g.,][]{Huang:2018}, especially those with short orbital periods.

Here, we report the discovery of one such system: \thisstar{}.  It is only the third star
known to have a transiting giant planet with an orbital period
shorter than 10 days as well
as a second transiting planet.  The host star is brighter than the host stars of
the previously known systems, especially at infrared wavelengths, which should
provide good opportunities to study this type of system in detail. The newly
discovered hot Jupiter also has a somewhat longer period (8.4 days) than 
WASP-47 (4.2 days) and Kepler-730 (6.5 days).
Thus, \thisstar{} may serve as a bridge connecting WASP-47 and Kepler-730 to longer-period giant planets.   

Section~\S\ref{sec:data} of this \emph{Letter} presents the \TESS{} photometric data, as well as the follow-up observations that validated both planet detections
and led to the measurement of the mass of the hot Jupiter.
Section~\S\ref{sec:analysis} describes our methods for determining the system parameters.
Section~\S\ref{sec:discusion} discusses the dynamical interactions between the two planets, as well as the possible implications this system will have on our understanding of hot Jupiter formation.

\section{Observations and Data Reduction}
\label{sec:data}

\subsection{\TESS\ photometry}

\target{} (TIC 254113311; \cite{Stassun:2019}) was observed by \TESS{} on CCD 2 of Camera 1 between June 19 and July 18, 2019, in the 13th and final sector of the survey of the southern ecliptic hemisphere. The star had not been pre-selected for 2-minute time sampling, and hence
the only available data are from the Full Frame Images (FFIs) with 30-minute sampling. We reduced the data using the Quick Look pipeline of \cite{Huang:2019}. Two sequences of transit signals were detected: \thisplanetb{}, with $P_\mathrm{b}=\,$4.07\,days and a signal-to-noise
ratio S/N of 24.2; and \thisplanetc{}, with $P_\mathrm{c}=\,$8.35\,days and\, S/N~$=78.2$.
Both signals passed the standard vetting tests employed by the \TESS{} Science Office, and the system was announced to the community as a \TESS{} Object of Interest (TOI).  

In an attempt to improve on the light curves produced by the automatic data reduction
pipeline, we performed multi-aperture photometry of the publicly available FFIs
that had been calibrated by the Science Processing Operation Center \citep{Jenkins:2016} (accessed via TESSCut\footnote{\url{https://mast.stsci.edu/tesscut/}.}).
Best results were obtained with a $3\times 3$ pixel square aperture centered on
the star. We omitted the data that were obtained at the beginning of the first spacecraft
orbit (BJD 2458653.93 to 2458657.72) because the data quality was compromised
by scattered moonlight.

The standard deviation of the time series of quaternions that the \TESS{} spacecraft
uses for attitude control has been shown to be correlated with systematic effects
in \TESS{} photometry. Therefore, we decorrelated the \thisstar{} light curve against the standard deviation of the Q1, Q2, and Q3 quaternion time series within each exposure, using a least squares technique. During this procedure, we excluded the data obtained during transits. We also iterated several times, removing 3$\sigma$ outliers from the fit until convergence \citep{Vanderburg:2019}. This process did not
remove a longer-term trend that was evident, but that is irrelevant for transit
analysis. We modeled this slower variability by adding a 4th-order polynomial to
the least-squares fit. Unlike \citet{Vanderburg:2019}, we did not perform
high-pass-filtering of the quaternion time series before the decorrelation.
Finally, we fitted a basis spline to the light curve to high-pass-filter any remaining long time scale variability (excluding transits and iteratively removing outliers, see \citealt{vj14}). 
\begin{figure*}
    \centering
    \includegraphics[width=0.9\linewidth]{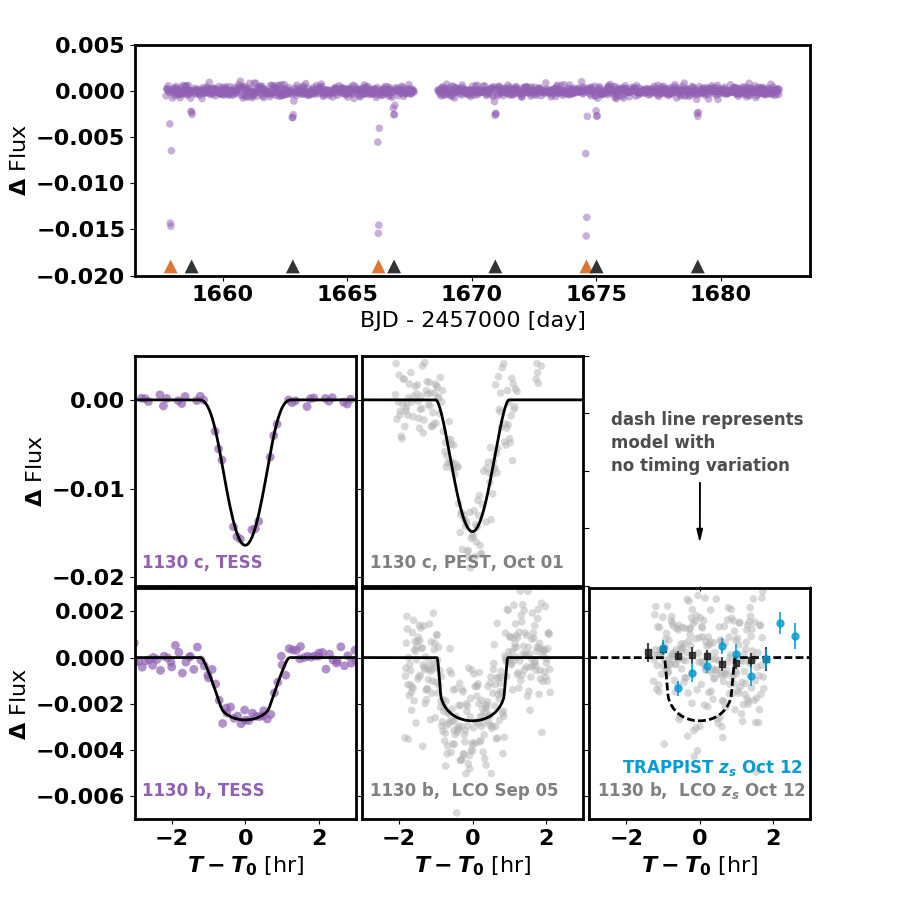}
    \caption{The light curve of \thisstar. The top panel shows the detrended discovery light curve from TESS, in units of fractional deviations from the out-of-transit level. The triangles along the time axis mark the times of transits of the two planets. The middle panels show the phase-folded \TESS\ light curve of the hot Jupiter as well as the follow-up light curve (gray points) from PEST obtained on UT 2019-Oct-01. The bottom panels show the phase-folded \TESS{} light curve of the inner planet as well
    as follow-up light curves from the Las Cumbres Observatory. The middle bottom panel
    shows $z_s$-band data from the Siding Spring Observatory node on 2019-Sep-05. The rightmost bottom panel shows the {\it non}-detection of the inner
    planet based on $z_s$-band observations from the
    Cerro Tololo node on 2019-Oct-12.  The black squares represent time-averaged data.
    The blue points show time-averaged $z_s$-band data from TRAPPIST South.   
    The black lines in all of the panels represent the best-fitting model assuming
    strict periodicity of the transits.}
    \label{fig:LC}
\end{figure*}

\subsection{Ground-based time-series photometry}

We conducted ground based seeing limited time-series photometric follow-up observations of \thisstar\ as part of the \TESS{} Follow-up Observing Program (TFOP).  To schedule these observations,
we used the {\tt TESS Transit Finder}, a customized version of the {\tt Tapir} software package \citep{Jensen:2013}. Observations were made with the Las Cumbres Observatory Global Telescope (LCOGT; \cite{Brown:2013}\footnote{https://lco.global}) network, the Perth Exoplanet Survey Telescope (PEST) in Australia, and the TRAPPIST-South telescope in Chile
\citep{jehin:2011,gillon:2013}.

A full transit of the inner planet \thisplanetb{} was observed in Pan-STARSS $z_s$ band on UT 2019-Sep-05 using a 1.0\,m telescope at the LCOGT Siding Spring Observatory (SSO) node.
The images from this observation and the other LCOGT observations
were calibrated with the standard BANZAI pipeline and light curves
were extracted using {\tt AstroImageJ} ({\tt AIJ}; \citealt{Collins:2017}).
An aperture radius of 2\arcsec\ was employed, which excluded most of the flux from a fainter star 4\arcsec\ away to the southeast ($\Delta T_{\rm mag} =6.9$).
The transit signal was clearly detected, with a duration and depth matching
the \TESS{} signal, thereby ruling out the
faint star as the source of the signal.
The bottom row of Figure~\ref{fig:LC} shows the light curve prepared with a 6\arcsec\ aperture, which gave a
higher signal-to-noise ratio than the 2\arcsec\ aperture.
Based on the star catalog from Gaia Data Release 2 \citep{Evans2018,GaiaDr2},
there are two other stars within 20\arcsec\ of \thisstar, but they are both too faint ($\Delta T_{\rm mag} = 8.6$ and 9.6) to be the source of the
\TESS{} signals.

We also observed one full transit of the hot Jupiter \thisplanetc{} in the Rc-band
with PEST on UT 2019-Oct-01.
PEST is a 12 {\arcsec} Meade LX200 SCT Schmidt--Cassegrain telescope
equipped with a SBIG ST-8XME camera located in a suburb of Perth, Australia.
A custom pipeline based on {\tt C-Munipack}\footnote{http://c-munipack.sourceforge.net} was used to calibrate the images and extract the differential time-series photometry. The transiting event was detected using
a 7.4\arcsec\ aperture centered on the target star.

We tried to observe another transit of \thisplanetb{} in both the B and $z_s$ bands on UT 2019-Oct-12 using a 1.0\,m telescope at the Cerro Tololo Interamerican Observatory
(CTIO) node of the LCOGT network. However, no transit signal was detected
within the 3-hour span of the observations, which had been timed to coincide with the
predicted time of transit.  The prediction was based on the \TESS{} data and
the assumption of a strictly periodic orbit. 
We began observing half an hour prior to the predicted ingress time, and ended one hour after the predicted
egress time. 
To make sure the transit
could have been detected, we injected a transit signal with the appropriate
characteristics into the LCOGT $z_s$-band light curve at the predicted epoch, which
made clear that the signal could have been detected
at the 10$\sigma$ level or higher. The data also show no evidence of an ingress or egress. Using the Bayesian information criterion comparing a transit model with the transit shape constrainted by the TESS data and a flat straight line model representing the scenario of no transit, we can confidently rule out that the center of transit happened inside the LCO observation baseline.  

Moreover, the TRAPPIST-South telescope at La Silla was also used to observe the same transit in the Sloan $z^\prime$-band.  No transit was detected.
Although the data are noisier than the LCO data, it is very likely that the transit
would have been detected if it had occurred on schedule without any timing deviations.
The non-detection is consistent with various scenarios in which the Neptune experiences large transit timing variations.

\subsection{Adaptive-optics images}

\begin{figure}
    \centering
    \includegraphics[width=0.9\linewidth]{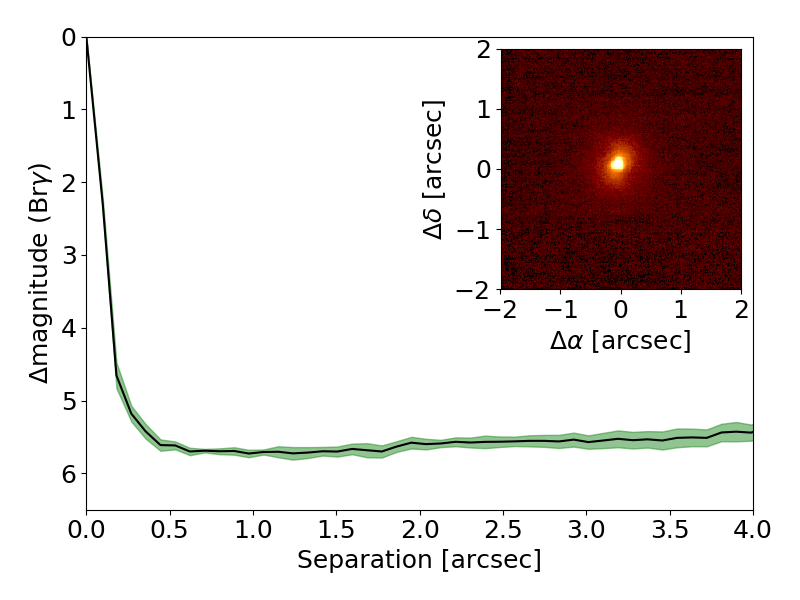}
    \caption{Br$\gamma$-band adaptive-optics image from VLT NaCo (inset), and the resulting sensitivity to visual companions as a function of angular separation. No companions
    were detected within the field of view.
    }
    \label{fig:AO}
\end{figure}

Adaptive-optics (AO) images were collected on UT 2019-Sep-14 using Unit Telescope 4 of the Very Large Telescopes (VLTs) equipped with the Naos Conica (NaCo) instrument.
We collected nine 20-second exposures with a Br$\gamma$ filter. The telescope
pointing was dithered by 2\arcsec\ in between exposures. Data reduction
followed standard procedures using custom IDL codes:
we removed bad pixels, flat-fielded the data, subtracted a sky background constructed from the dithered science frames, aligned the images, and co-added the data to obtain
the final image. The sensitivity to faint companions was determined by injecting scaled
point-spread-functions (PSFs) at a variety of position angles and separations.
The scaling was adjusted until the injected point sources could
be detected with 5$\sigma$ confidence. No companions were detected down
to a contrast of 5.7\,mag at 1\arcsec. Figure \ref{fig:AO} shows the sensitivity curve as a function of angular separation, along with a small image of the immediate environment of
\thisstar{}.

\subsection{Radial velocities}


We obtained a series of spectra of \thisstar{} using the CHIRON facility \citep{2013PASP..125.1336T} to monitor the star's radial-velocity variations
and thereby measure or constrain the masses of the planets.
CHIRON is a high-resolution spectrograph on the SMARTS 1.5\,m telescope at CTIO.
Light is delivered to the spectrograph via an image slicer and a fiber bundle, with a resolving power of $80{,}000$ over the wavelength range from 4100 to $8700$\,\AA{}. A total of 21 spectra were obtained between UT 2019-Aug-30 and UT 2019-Oct-17. 
There are no stars in the Gaia DR2 catalog that would have fallen within the CHIRON fiber (2.7\arcsec in radius) that could contaminate the RVs. 

The radial velocities were measured from the extracted spectra by modeling the least-squares deconvolution line profiles \citep{1997MNRAS.291..658D}. Table~\ref{tab:rv} gives the results.

\begin{figure*}[ht!]
    \centering
    \includegraphics[width=0.9\linewidth]{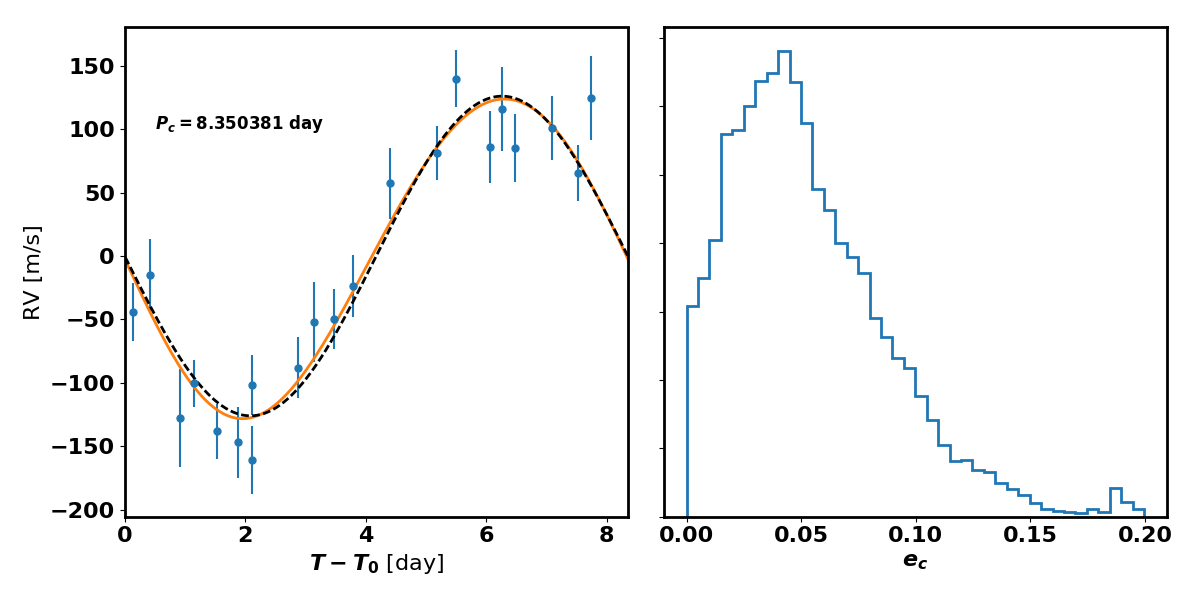}
    \caption{The left panel shows the relative radial-velocity orbit of \thisplanetc{} based on CHIRON data. The plotted error bars include the ``jitter'' term described
    in Section~\ref{sec:analysis}.
    The orange line is the best-fitting model.  The black dashed line represents a circular orbit with the same semi-amplitude. The right panel shows the posterior
    probability distribution for the orbital eccentricity.}
    \label{fig:RV}
\end{figure*}

\section{Analysis}
\label{sec:analysis}

\subsection{Stellar parameters} 
\label{sec:star}


We determined the basic stellar parameters by fitting the observed
spectral energy distribution (SED) \footnote{We also derived the best-fitting stellar parameters from the average CHIRON spectra, yielding $T_\mathrm{eff} = 4545\pm14$\,K, $\log g_\star = 4.60 \pm0.038$ dex, $\mathrm{[m/H]} = -0.105\pm0.063$ dex, and $v\sin I_\star = 4\,\mathrm{km\,s}^{-1}$. However, because the library is not well calibrated for low-mass stars, we did not rely on these CHIRON-based parameters in the subsequent analysis.}.
We compared the available broadband photometry
with the M/K dwarf spectral templates of \citet{Gaidos2014} and \citet{Kesseli2017}.
The details of this SED-fitting procedure were described by \citet{Mann2015b} and
are summarized here. For the photometry, we consulted the star catalogs from
the Two-Micron All-Sky Survey \citep[2MASS,][]{Skrutskie2006}, the Wide-field Infrared Survey Explorer \citep[WISE,][]{Wright2010}, the Gaia DR2 \citep{Evans2018,GaiaDr2}, the AAVSO All-Sky Photometric Survey \citep[APASS,][]{Henden2012}, and Tycho-2 \citep{Hog2000}. We compared the observed magnitudes to the synthetic magnitudes computed from each template
spectrum, using Phoenix BT-SETTL models \citep{Allard2011} to fill in the gaps in the spectra. We did not account for reddening or extinction, because the star is within the local bubble where these effects should be negligible. The resulting
parameters are $\teff = 4250\pm67$\,K, bolometric flux $= (1.42\pm0.05)\times 10^{-9}$\,\ergscmsq, $L_\star=0.150\pm0.006\,L_\odot$, and $R_\star=0.714\pm0.029\,R_\odot$. The best-fitting template and model combination gave a minimum reduced
chi-squared of 0.8, indicating a good fit.
These results are consistent with the standard stellar SED fitting method using the NextGen stellar atmosphere models \citep{Stassun:2018, Stassun:2019}, which gave $T_{\rm eff} = 4300 \pm 100 K$, and $R_\star = 0.692 \pm 0.032 R_\odot$. 

The SED fit strongly favors a metal-rich composition. All of the templates
with a solar or sub-solar metallicity gave $\chi^2_\nu\,>\,3$. 
The Gaia data also reveals that
the $M_G$ absolute magnitude of \target\ places it within the
brightest 10\% of stars with the same $BP-RP$ color.
Since late-K dwarfs do not evolve significantly over the lifetime of the Universe, this high position in the color-magnitude diagram is best explained by a high metallicity.
(The possibility of an unresolved stellar companion is ruled out
by the adaptive-optics imaging presented above.)
Based on the expected distribution of metallicities in the Solar neighborhood, we infer that \target\ has a metal content [M/H]$\,>\,0.2$. 

We estimated $M_\star$ using the empirical relation between $M_{K_S}$ and mass from \citep{Mann2019a}\footnote{\href{https://github.com/awmann/M_-M_K-}{https://github.com/awmann/M\_-M\_K-}}. This relation was calibrated using dynamical masses of K and M dwarf binaries.  The result is $M_\star= 0.671\pm0.018\,M_\odot$.

\subsection{Global Modeling}

We performed a joint analysis of the \TESS{} transit light curve,  the 21 radial
velocities from CHIRON, and the ground-based follow-up light curves excluding the Oct 12th observations. 
We restricted the orbital
eccentricity of \thisplanetc{} to be smaller than 0.2. 
Numerical integrations showed that the system would not be stable for more than $10^6$ orbits if the eccentricity were any larger.
We also allowed for a radial-velocity ``jitter'' term, which was added in quadrature to the nominal uncertainties to account for unmodeled systematic and astrophysical effects. 
We did not include the effects of \thisplanetb{} in the radial velocity model,
because the expected radial-velocity amplitude is beneath the 10~m\,s$^{-1}$ level. 

We assumed that the stellar limb-darkening follows a quadratic law and used the formulas of \citet{MandelAgol:2002} as implemented by \citet{Kreidberg(2015)} while modeling the transit light curves. We set priors on the limb-darkening coefficients using the LDTk model implemented by \citet{Parviainen2015} based on a library of PHOENIX-generated specific intensity spectra by \citet{Husser2013}. The resulting limb-darkening coefficients are consistent with the values tabulated by \citet{Claret:2018} and \citet{Claret:2012}. To account for the 30-minute averaging time of the \TESS\ data, the photometric model was computed with 1~min sampling and then averaged to 30~minutes \citep{Kipping2010}.

The mass and radius of the star were also adjustable parameters, with priors
based on the results presented in Section \ref{sec:star}.
Another constraint on these parameters came from
the implicit value of the stellar mean density $\rho_\star$ that arises
from the combination of $P$, $a/R_\star$, and $i$ \citep{SeagerMallenOrnelas2003,Winn2010}.
The likelihood function enforced agreement
with the measurements of $\rho_\star$ from the posterior determined by the SED modeling.

To determine the credible intervals for all the parameters,
we used the ``emcee'' Markov Chain Monte Carlo method of
\citet{ForemanMackey:2012}.
The results are given in Table~\ref{tab:stellar},
and the
best-fitting model is plotted in Figures~\ref{fig:LC} and
\ref{fig:RV}.
For a ``second opinion'' on the model parameters,
we used the EXOFASTv2 code \citep{Eastman:2013, Eastman:2019} to fit the
same data.  The results all agreed to within 0.5$\sigma$ or better.

The model assumed the transits to be strictly periodic, despite the evidence
for transit-timing variations presented earlier. We did not account for the Oct 12 observation in our global modeling.
For this reason, we caution that the uncertainties in the orbital
periods are likely larger than are reported in Table~\ref{tab:stellar}. 
This is especially true for the lower-mass planet \thisplanetb{}.
Further photometric observations are needed to get a better understanding
of the periods and the timing variations.    

\subsection{Confirmation of \thisplanetc}

The mass of \thisplanetc{} was found to be \tessfitMpc{} \mjup{}. 
The radius of the planet is not well constrained because the transit is grazing. However, based on the mass of the planet, we put a prior constraint on the radius of the planet to be less than 2 \rjup{}, and are able to determine the radius to be \tessfitRpc{} \rjup{}. The orbit of \thisplanetc{} appears to be slightly eccentric, $e=$\tessfitec{}. Modeling the CHIRON RVs alone would give a more eccentric solution for \thisplanetc{}, $e=0.074\pm0.023$. Future monitoring will refine the eccentricity constraint.    

\subsection{Validation of \thisplanetb}

The CHIRON data are not precise enough to reveal the radial-velocity signal of \thisplanetb{}. An upper limit on the mass of \thisplanetb{} was obtained by fitting a two planet model to the radial velocity data, using the posterior of the global modeling to constraint the period and epoch of both planets. We allow the semi-amplitude to be negative in the fit.  
The resulting 3$\sigma$ upper limit is $40$ \mearth.
Even though the radial-velocity signal was not detected, there is a 2$\sigma$ hint that the orbit of \thisplanetb{} is eccentric, based on the combination of the
transit duration, transit impact
parameter, and the observational constraints on the mean stellar density.

Without a radial-velocity detection, one must proceed with care to make sure
that the \TESS\ transit signals really arise from a planet around the target star,
and not an unresolved background eclipsing binary or other type of ``false positive.''
The transit signals seen by \thisplanetb{} in \TESS\ and LCOGT have a flat bottom,
in contrast to the V-shaped appearance of most
eclipsing binaries. 

A more quantitative argument can be made based on the ratio between the duration of ingress or egress and the duration of the flat-bottomed portion of the transit
\citep{SeagerMallenOrnelas2003}.
This ratio is observed to be $T_{12}/T_{13}=0.064\pm0.02$.
For an isolated star with an eclipsing companion, this ratio is equal to
the maximum possible radius ratio between the eclipsing object
and the star.  The corresponding
maximum flux deficit is the square of the radius ratio, giving a 2$\sigma$ upper limit
on the flux deficit of 0.007.
To produce such a signal, a blended stellar companion would need to be
within 1.23 magnitudes of \thisstar. The adaptive-optics image presented
in \S~\ref{sec:data} rules out such a companion beyond 1\arcsec\ (a projected separation of $\mysim$58 AU). Based on the lack of any long-term trend in the CHIRON radial-velocity
data, we are also able to place a 3$\sigma$ upper limit of
$0.318\,\msun$ ($\Delta$\,mag~$\lesssim 2.6$)
on any bound companion within 4 AU.

We used \texttt{vespa} \citep{Morton:2015} to evaluate the probability of any remaining
false positive scenarios involving eclipsing binaries.
Using the \TESS\ light curve of \thisplanetb{} and the constraints from spectroscopy and imaging, \texttt{vespa} returns a false positive probability of $FPP< 10^{-6}$.
Thus, we consider \thisplanetb{} to be a validated planet.
Section \ref{sec:dynamic} presents further evidence that this
planet orbits the same star as \thisplanetc{},
based on the tentative detection of transit timing variations.

\begin{deluxetable}{crr}
\tablewidth{0pc}
\tabletypesize{\scriptsize}
\tablecaption{
    Radial velocity for \target
    \label{tab:rv}
}

\tablehead{
    \colhead{time} &
    \colhead{RV [${\mathrm km}\,{\mathrm s}^{-1}$]}                     &
    \colhead{error [${\mathrm km}\,{\mathrm s}^{-1}$]}    
}
\startdata
2458725.63420 &  -9.652 & 0.037 \\
2458734.59335 & -9.662 & 0.018 \\
2458738.55132 & -9.384 & 0.020 \\
2458739.53575 & -9.439 & 0.024 \\
2458740.57420 & -9.459 & 0.019 \\
2458741.54135 & -9.568 & 0.020 \\
2458742.55255 & -9.624 & 0.015 \\
2458743.52740 & -9.685 & 0.025 \\
2458744.55082 & -9.576 & 0.030 \\
2458746.59182 & -9.443 & 0.018 \\
2458747.66486 & -9.408 & 0.031 \\
2458751.63354 & -9.671 & 0.026 \\
2458752.63038 & -9.612 & 0.021 \\
2458753.54111 & -9.547 & 0.022 \\
2458757.50133 & -9.400 & 0.031 \\
2458758.52657 & -9.539 & 0.026 \\
2458761.58060 & -9.574 & 0.021 \\
2458762.51420 & -9.467 & 0.025 \\
2458768.57449 & -9.626 & 0.021 \\
2458772.53023 & -9.438 & 0.026 \\
2458773.55102 & -9.423 & 0.022 \\
\enddata
\end{deluxetable}

\section{Discussion}
\label{sec:discusion}

\subsection{Dynamical Constraints} 

Dynamical simulations were conducted, with the hope
of improving our knowledge of the system parameters by requiring
that they be consistent with long-term stability.
We also wanted to see if transit-timing variations due
to planet-planet interactions
could plausibly be large enough
to explain the ``missing transit'' on October 12, 2019.
\label{sec:dynamic}

\subsubsection{System Stability}

We performed three suites of simulations using Mercury6 \citep{mercury6}. The first two suites were composed of 100 simulations each.
The initial conditions for each simulation
were selected from a 
randomly chosen link in the posterior produced by the analysis described
in Section \ref{sec:analysis}.
However, since the mass, eccentricity, and argument of pericenter $\omega$ of the inner planet are poorly constrained,
the initial values of those parameters were handled differently.
The mass of the planet was set equal to that of Neptune.
In the first suite of simulations, we set the initial eccentricity
equal to zero.  
In the second suite, we drew $e$ and $\omega$ from uniform distributions
with ranges of 0.0--0.3 and 0--360$^\circ$, respectively.

We used a time-step of 20 minutes to integrate
the equations of motion for $10^{5}$ years,
used the hybrid symplectic and Bulirsch-Stoer integrator,
and enforced energy conservation to within one part in $10^{8}$ or better. 

In both suites of simulations, the vast majority of initial
conditions led to stable configurations, i.e.,
they did not experience orbit crossings, collisions, or ejections during the simulation time.
In the first suite, all 100 trials were consistent with stability.
In the second suite,  were stable.
The four unstable trials involved some of the highest initial
eccentricities for both planets. These experiments suggest that if \thisplanetb\ has a low eccentricity, essentially the full range
of system parameters consistent with the data
are also consistent with dynamical stability.
A moderate eccentricity for the inner planet is also generally
consistent with long-term stability.

In the stable configurations, the planetary eccentricities oscillate. For TOI-1130~b, the forced eccentricity is the most important component. The largest value obtained in the dynamically stable trials of either
suite was about 0.30. The typical value of upper envelope of the eccentricity oscillations was closer to 0.17.
The relative contributions of the free and forced eccentricities can be determined better through future observations of the phase of the TTVs.

The third suite of simulations, composed of 500 integrations, was intended to study the planetary eccentricities in more detail.
We tested a large range of possible eccentricities for both planets (while randomizing $\omega$). The inner planet was assumed to have the same mass as Neptune. The outer planet's mass was drawn from the posterior, along with all of the other system parameters.
Dynamical stability was seen in all the trials for which
the eccentricities obeyed the relation $e_b + 2e_c < 0.4$.
When this inequality was violated, instability was more likely.
If $e_b$ rose above 0.4 or 0.5, the system was nearly always unstable.

\subsubsection{Transit Timing Variations}
\label{subsec:ttvs}

As described in \S~\ref{sec:data}, the two attempts to observe
the transit of October 12, 2019 resulted in flat light curves,
ruling out the occurrence of a transit at the predicted time.
Could this plausibly be due to a large transit-timing variation
caused by planet-planet interactions?

The ratio between the orbital periods of the
two planets is within 2.5\% of 2:1, implying that the system
is close to resonance. This condition usually results in large TTVs.
Based on the current best estimates of the orbital periods,
the super-period of the expected TTVs, computed using the analytic theory of \cite{Lithwick2012}, is between about 156 and 156.5 days.
Inflating the error on each orbital period to 1.5 minutes, however, increases the uncertainty on the super period by of a factor of 16 to about 8 days.
Although the super-period is fairly well constrained, the expected amplitude
of the timing variations is poorly constrained. The unknown
mass of the inner planet leads to estimates for the
TTV amplitudes ranging from seconds to hours.

\begin{figure}
    \centering
    \includegraphics[width=3.4in]{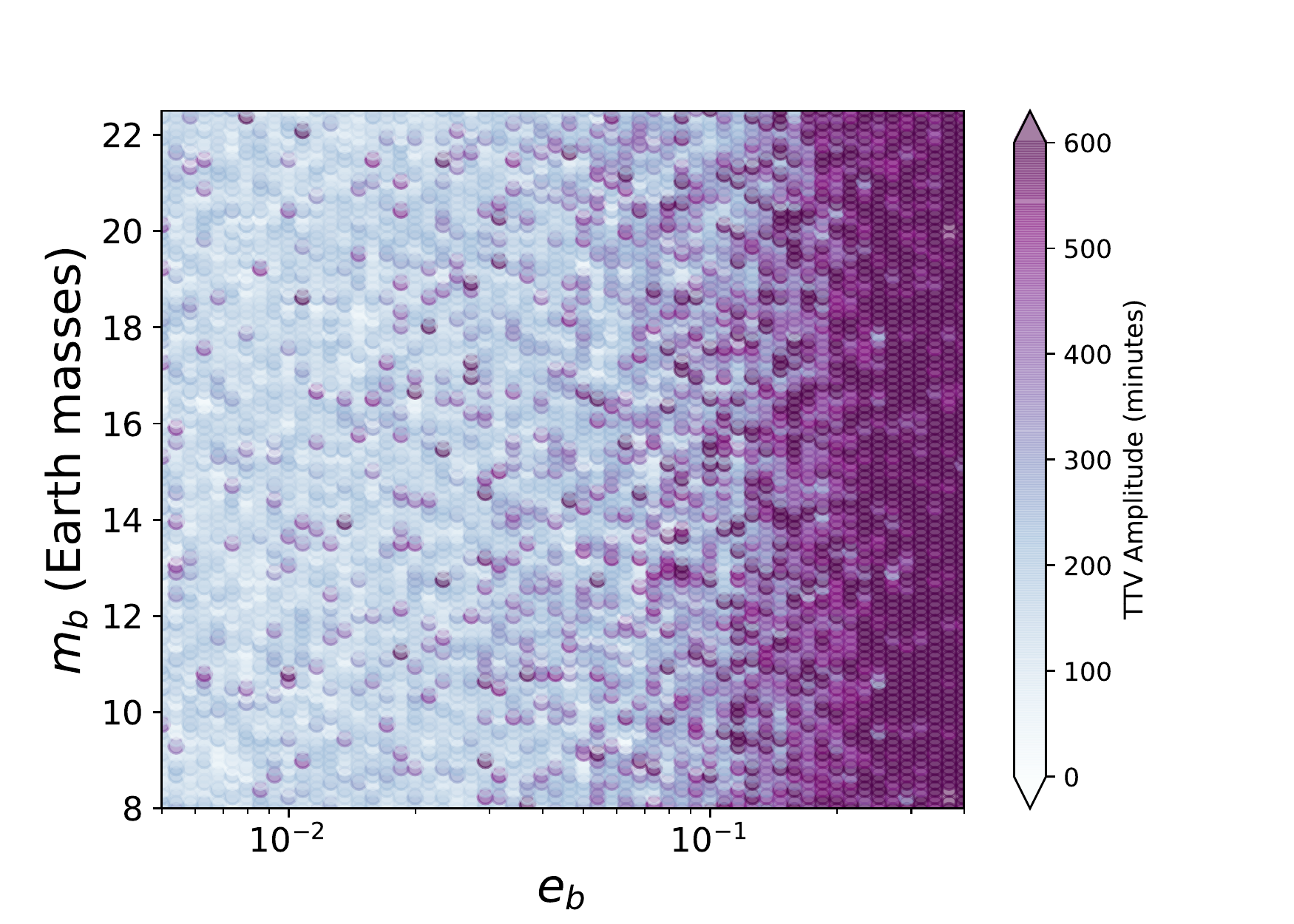}
    \caption{Monte Carlo exploration of the theoretical TTV amplitude
    as a function of the mass and eccentricity of TOI-1130\,b. Each point corresponds to one link drawn from the posterior. The color encodes
    the TTV amplitude. For eccentricities exceeding about 0.01, the typical TTV amplitude is on the order of hours, which is large enough to explain
    the non-detection of the October 12 transit.}
    \label{fig:ttv_theory}
\end{figure}

Figure \ref{fig:ttv_theory} shows the dependence of the TTV amplitude on the mass and eccentricity of the inner planet. The TTV amplitude was computed using TTVFast \citep{Deck:2014}.
In these Monte Carlo trials, the stellar parameters and those of the outer planet's orbital elements were drawn from the posterior, while the inner planet's mass and eccentricity were sampled uniformly between the limits shown on the plot. The argument of pericenter was drawn
randomly from a uniform distribution. To explain the October transit non-detection, we require a TTV amplitude of at least two hours. The majority of parameter space is expected to give TTV amplitudes at this level or above.  Thus, TTVs are indeed a plausible explanation.

\begin{figure*}
    \centering
    \includegraphics[width=0.9\linewidth]{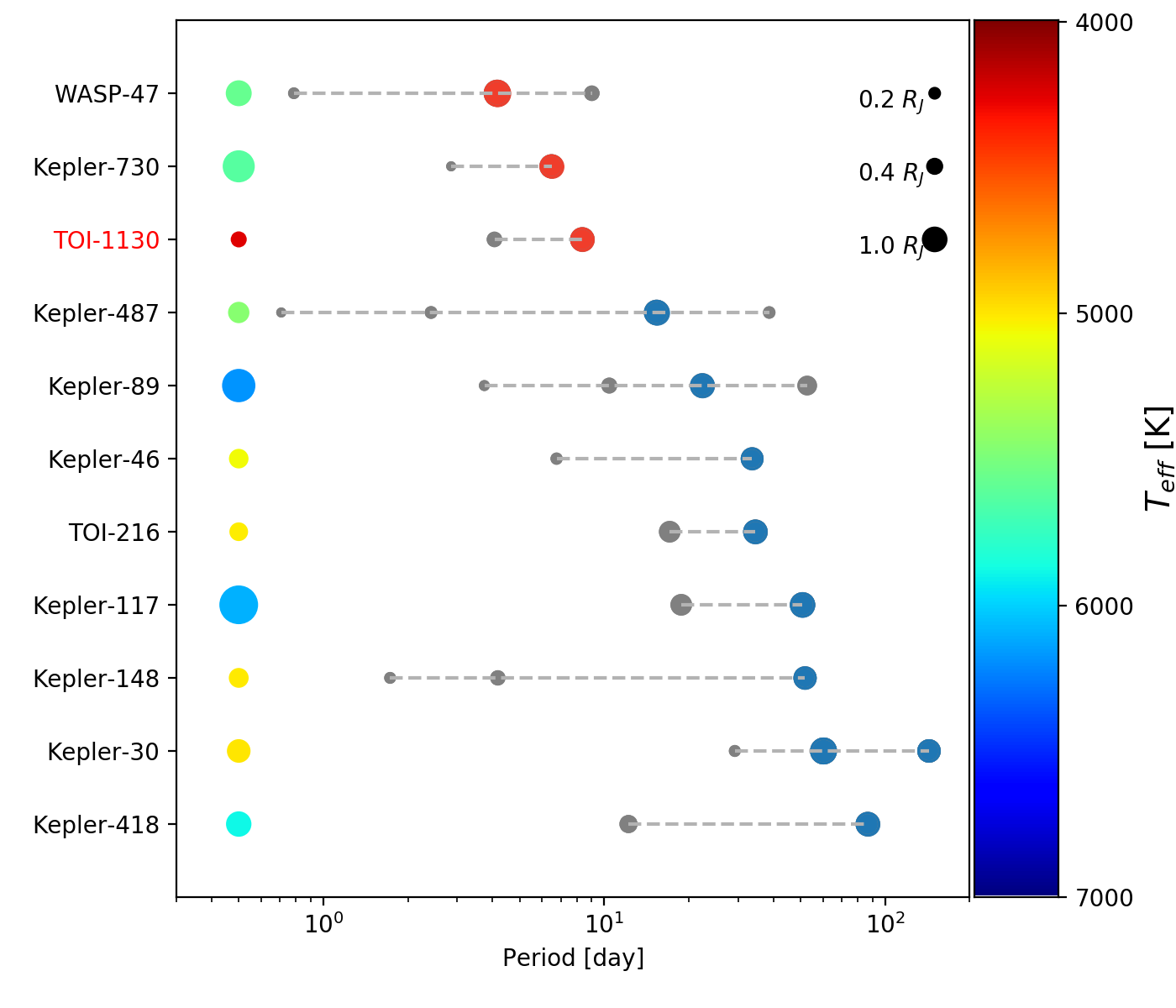}
    \caption{All confirmed planetary systems consisting of a transiting giant planet with period shorter than 100 days and inner transiting companions. Each horizontal line represents a planetary system. The giant planets with period smaller (larger) than 10 days are represented by red (blue) circles, and the small planets are represented by gray circles. The first circle in each line represent the host star, color coded with their effective temperature. The sizes of the circles are proportional to the radii of planets. }
    \label{fig:systems}
\end{figure*}

\subsection{TOI-1130's place in the hot Jupiter paradigm} 

Figure~\ref{fig:systems} illustrates the period distribution
of transiting giant planets with transiting inner companions.
The only three transiting hot Jupiters ($P<10$\,days) known to have
inner transiting companions are WASP-47\,b, Kepler-730\,b, and
TOI-1130\,c. Their orbital periods are 
4.2 days, 6.5 days, and 8.4 days, respectively.
Other giant planets with somewhat longer orbital periods --- ``warm Jupiters'' --- are more frequently found with inner companions \citep{Huang:2016}.

The apparently continuous period distribution of the giant
planets in Figure~\ref{fig:systems} suggests that the hot
Jupiters with inner companions are not so different from
the warm Jupiters with inner companions.  Perhaps both types
of systems are produced by the same process, and the hot
Jupiters with companions represent the tail of a statistical distribution of outcomes.  In that case, the more commonly
encountered ``lonely'' hot Jupiters (without close companions)
might have formed from a
different mechanism.

Comparison of Figure~\ref{fig:systems} with similar
figures that have been made for {\it Kepler} systems in general
\citep[see, e.g.,][]{Fabrycky:2014} suggest that
the systems with
giant planets and small inner companions resemble 
the closely packed and coplanar
{\it Kepler} multi-planet systems of super-Earths. 
The giant-planet systems simply have more extreme size and 
mass ratios between the planets.
Specifically, in Kepler multi-planet systems in general (which mostly contain sub-Neptune sized planets, the typical Hill spacing is 21.7$\pm$ 9.5 (e.g., \citet{Fang:2013, Weiss:2018}). The typical mutual Hill radii of the planets in Figure 5 is 16.8$\pm$9.6.
We speculate that all these close-orbiting multi-planet
systems originated from essentially the same process,
but in rare cases, one of the super-Earths
managed to exceed the threshold mass for runaway gas accretion
\citep{Lee:2014, Batygin:2015}.
Such rare cases may lead to the formation of the systems shown
in Figure~\ref{fig:systems}.

One reason why it would be interesting to further enlarge the sample of giant planets with small inner companions is to study
the distribution of period ratios, and the proximity to
resonances.  In all three cases of hot Jupiters with
inner companions, none of the known planets are in resonance.
Only a small fraction of the {\it Kepler} multi-planet systems are
in resonances \citep{Lissauer:2011}, while systems of multiple
wide-orbiting giant planets are frequently in resonance \citep{WinnFabrycky2015}. It would therefore be interesting
to know how frequently systems with giant planets and small inner
companions are in resonance. If these systems and the super-Earth systems both assembled via the same mechanism, then one might expect the period ratio distributions (including the occurrence of resonances) to be similar.  

\thisstar{} has a brighter host star than WASP-47 or Kepler-730,
which will facilitate follow-up opportunities to investigate the mystery the formation of these types of systems. For example, the expected Rossiter--McLaughlin amplitudes of the two planets (6\,${\mathrm m}\,{\mathrm s}^{-1}$ and 7 ${\mathrm m}\,{\mathrm s}^{-1}$)\footnote{Even though the transit of TOI 1130 c is significantly deeper than that of TOI 1130 b, we expect they will have similar Rossiter-McLaughlin amplitudes because of TOI 1130 c's high impact parameter.} are detectable with current facilities for a star as bright as \thisstar. These measurements can reveal the stellar obliquity and mutual inclination between the orbits, both of which are relevant
to the formation mechanism. Additionally, \thisstar{} is a K7 star, which is the smallest star known to host similar type of system architecture to-date. It is relatively bright at near-infrared wavelengths
($K_s = 8.351$), making the planet a good target for transit spectroscopy
to study planetary atmospheres.  Specifically, the
atmospheric signal of \thisplanetc{} is probably
detectable with the {\it Hubble
Space Telescope.} 
Comparisons between its atmosphere and that
of the other hot and warm Jupiters may help us understand its origin.

The discovery of \thisstar{} illustrates \TESS's power to find systems with rare architectures. With a large amount of \TESS{} data still unexplored, we can expect more systems such as \thisstar{},
along with better knowledge of the frequencies of different
types of hot Jupiter systems.

\acknowledgments

We thank the \TESS\ Mission team and Follow-up Working Group for the valuable dataset. 
This paper includes data collected by the \TESS\ mission, which are publicly available from the Mikulski Archive for Space Telescopes
(MAST). Funding for the \TESS\ mission is provided by NASA's Science Mission directorate.
CXH and MNG acknowledge support from MIT's Kavli Institute as Torres postdoctoral fellows.
AV's work was performed under contract with the California Institute of Technology / Jet Propulsion Laboratory funded by NASA through the Sagan Fellowship Program executed by the NASA Exoplanet Science Institute. JJL's work was supported by the TESS GI grant G011108. 
JNW's work was partly supported by the Heising-Simons Foundation.
Resources supporting this work were provided by the NASA High-End Computing (HEC) Program through the NASA Advanced Supercomputing (NAS) Division at Ames Research Center for the production of the SPOC data products. This work is based in part on observations collected at the European Organisation for Astronomical Research in the Southern Hemisphere under ESO program P103.C-0449.
MF, IG and CMP gratefully acknowledge the support of the  Swedish National Space Agency (DNR 163/16 and 174/18). The research leading to these results has received funding from the European Research Council under the European Union's Seventh Framework Programme (FP/2007-2013) ERC Grant Agreement n$^\circ$ 336480, from the ARC grant for Concerted Research Actions, financed by the Wallonia-Brussels Federation. TRAPPIST is funded by the Belgian Fund for Scientific Research (Fond National de la Recherche Scientifique, FNRS) under the grant FRFC 2.5.594.09.F, with the participation of the Swiss National Science Fundation (SNF). M.G. and E.J. are FNRS Senior Research Associates. This work makes use of observations from the LCOGT network.
 
\software{}We made use of the Python programming language \citep{Rossum1995} 
and the open-source Python packages
\textsc{numpy} \citep{vanderWalt2011}, 
\textsc{emcee} \citep{Foreman-Mackey2013}, 
\textsc{batman} \citep{Kreidberg(2015)}
{\scshape rebound} \citep{rebound}.
We also used Mercury \citep{mercury6} and AstroImageJ \citep{Collins:2017}.

\textit{Facilities:}

TESS, CHIRON, LCOGT, PEST, TRAPPIST-South, VLT

\begin{deluxetable*}{lcr}
\tablewidth{0pc}
\tabletypesize{\tiny}
\tablecaption{
    System Parameters for \target
    \label{tab:stellar}
}
\tablehead{
\colhead{{\bf Parameters}} & \colhead{{\bf Values}} & 
\colhead{{\bf Comments}}}
\startdata
\noalign{\vskip -5pt}
\sidehead{\bf Catalog Information}
~~~~R.A. (h:m:s)                      &   19:05:30.24  & \gaia{} DR2\\
~~~~Dec. (d:m:s)                      &  $-41$:26:15.49   & \gaia{} DR2\\
~~~~Epoch							  &  2015.5         & \gaia{} DR2 \\
~~~~Parallax (mas)                    & $17.13\pm0.049$ & \gaia{} DR2\\
~~~~$\mu_{\mathrm{ra}}$ (mas yr$^{-1}$)        & $12.54\pm0.088$  & \gaia{} DR2 \\
~~~~$\mu_{\mathrm{dec}}$ (mas yr$^{-1}$)       & $-27.18\pm0.071$ & \gaia{} DR2\\
~~~~\gaia{} DR2 ID                       &  6715688452614516736  &  \\
~~~~Tycho ID                             & TYC 7925-02200-1   & \\
~~~~TIC ID                            &  254113311 & \\
~~~~TOI ID                            &  1130    & \\
\noalign{\vskip -5pt}
\sidehead{\bf Photometric properties}
~~~~$B$ (mag)\dotfill               & 12.632 & APASS    \\
~~~~$V$ (mag)\dotfill               & 11.368  &  APASS    \\    
~~~~\TESS{} (mag)\dotfill            &  10.143 & TIC V8         \\
~~~~\gaia{} (mag)\dotfill            & 10.902 & \gaia{} DR2               \\
~~~~\gaia{}$_r$ (mag)\dotfill          &  10.092
 & \gaia{} DR2                 \\
~~~~\gaia{}$_b$ (mag)\dotfill          & 11.653 & \gaia{} DR2                 \\
~~~~$J$ (mag)\dotfill               & $9.055\pm0.023$ & 2MASS           \\
~~~~$H$ (mag)\dotfill               & $8.493\pm0.059$ & 2MASS           \\
~~~~$K_s$ (mag)\dotfill             & $8.351\pm0.033$  & 2MASS           \\
\noalign{\vskip -5pt}
\sidehead{\bf Stellar properties}
~~~~$\teffstar$ (K)\dotfill        & $4250\pm67$   & SED  \\
~~~~$\loggstar$ (cgs)\dotfill       & \tessfitloggstar{}  &   this work \\
~~~~[Fe/H] (dex)\dotfill       &  $>0.2$ & SED     \\
~~~~$v \sin i$ (\kms)\dotfill            &  $4.0\pm0.5$ & SPC \\ 
~~~~$\mstar$ ($\msun$)\dotfill      & \tessfitmstar{}  & this work \\
~~~~$\rstar$ ($\rsun$)\dotfill      & \tessfitrstar{} &  this work         \\
~~~~$\lstar$ ($\lsun$)\dotfill      & \tessfitlumstar{} & this work        \\
~~~~Age (Gyr)\dotfill               &  $8.2_{-4.9}^{+3.8}$  &    this work  \\
~~~~Distance (pc)\dotfill           &  $58.26\pm0.17$ & \gaia{} DR2\\
~~~~$\rhostar$ (\gcmc)\dotfill & \tessfitrhostar{} & this work  \\
~~~~$u_{1, \rm tess}$ \dotfill & 0.49 $\pm$ 0.05 & this work \\
~~~~$u_{2, \rm tess}$ \dotfill & 0.06 $\pm$ 0.05& this work\\
~~~~$u_{1, \rm z_s}$ \dotfill & 0.46$\pm$  0.05& this work \\
~~~~$u_{2, \rm z_s}$ \dotfill & 0.13$\pm$ 0.05&this work  \\
~~~~$u_{1, \rm R_c}$ \dotfill & 0.72 $\pm$ 0.05	 & this work\\
~~~~$u_{2, \rm R_c}$ \dotfill & 0.06 $\pm$ 0.03 & this work\\
\noalign{\vskip -5pt}
\sidehead{\bf Additional RV parameters}
~~~~ $\gamma$(\kms) \dotfill & \chironfitgamma{} &  \\
~~~ jitter (\kms)  \dotfill & \chironfitjit{} & \\ 
\multicolumn{1}{l}{{\bf Planet Parameters}}& \multicolumn{1}{c}{ b} & \multicolumn{1}{c}{c} \\
~~~$P$ (days)             \dotfill    &     \multicolumn{1}{c}{\tessfitPb{}}    &       \multicolumn{1}{c}{\tessfitPc{}}   \\
~~~$T_{c}$ (${\rm BJD} $)     \dotfill    &  \multicolumn{1}{c}{\tessfitTcb{}}       & \multicolumn{1}{c}{\tessfitTcc{}}    \\
~~~$K$ (\kms)     \dotfill    &        &  \multicolumn{1}{c}{\tessfitKc{}}   \\
~~~$\sqrt{e} \cos \omega$      \dotfill    &  \multicolumn{1}{c}{\tessfitecoswb{}}   &   \multicolumn{1}{c}{\tessfitecoswc{}}    \\
~~~$\sqrt{e} \sin \omega$      \dotfill    &  \multicolumn{1}{c}{\tessfitesinwb{}}   & \multicolumn{1}{c}{\tessfitesinwc{}}    \\
~~~$e$ \dotfill & \multicolumn{1}{c}{\tessfiteb{}} & \multicolumn{1}{c}{\tessfitec{}} \\
~~~$\omega$ \dotfill & - & \multicolumn{1}{c}{\tessfitwc{}} \\
~~~$T_{14}$ (hrs)  \dotfill    & \multicolumn{1}{c}{\tessfitTdurb{}}   &  \multicolumn{1}{c}{\tessfitTdurc{}}     \\
~~~$\arstar$              \dotfill    &  \multicolumn{1}{c}{\tessfitaorb{}}   &  \multicolumn{1}{c}{\tessfitaorc{}} \\
~~~$\rpl/\rstar$          \dotfill    & \multicolumn{1}{c}{\tessfitRratiob{}}    &  \multicolumn{1}{c}{\tessfitRratioc{}}   \\
~~~$b \equiv a \cos i/\rstar$
                          \dotfill    &   \multicolumn{1}{c}{\tessfitbb{}}   & \multicolumn{1}{c}{\tessfitbc{}}   \\
~~~$i_c$ (deg)              \dotfill    &  \multicolumn{1}{c}{\tessfitincb{}}  &    \multicolumn{1}{c}{\tessfitincc{}} \\
~~~$\mpl$      \dotfill    &  -     &  \multicolumn{1}{c}{\tessfitMpc{}\mjup}\\
~~~$\rpl$       \dotfill    &  \multicolumn{1}{c}{\tessfitRpb{}\rearth }  &\multicolumn{1}{c}{\tessfitRpc{}\rjup } \\
~~~$\rhopl$ (\gcmc)       \dotfill    & -
& \multicolumn{1}{c}{\tessfitrhoc{}} \\
~~~$a$ (AU)               \dotfill    &  \multicolumn{1}{c}{\tessfitsemib{}}    & \multicolumn{1}{c}{\tessfitsemic{}}  \\
~~~$T_{\rm eq}$ (K)        \dotfill   & \multicolumn{1}{c}{\tessfitTeqb{}}    & \multicolumn{1}{c}{\tessfitTeqc{}}  \\  
\enddata
\end{deluxetable*}

\bibliographystyle{apj}

\end{document}